\documentclass[aps,reprint,10pt,showkeys, superscriptaddress,showpacs]{revtex4-2}
\usepackage{multirow}
\usepackage{enumitem}
\usepackage{amsfonts} 
\usepackage{amsmath}
\usepackage{amssymb}
\usepackage{graphicx}
\usepackage{subfigure}
\usepackage{color}
\usepackage{enumerate}
\usepackage[]{natbib}
\usepackage{sidecap}
\usepackage{soul}
\usepackage{cancel}
\usepackage[normalem]{ulem}
\usepackage{dcolumn}%
\usepackage{bm}%

\newcommand*\xbar[1]{%
  \hbox{%
    \vbox{%
      \hrule height 0.5pt 
      \kern0.5ex
      \hbox{%
        \kern-0.1em
        \ensuremath{#1}%
        \kern-0.1em
      }%
    }%
  }%
} 
\begin{document}
\title{Ion-beam driven dust-cyclotron and dust-lower-hybrid instabilities in nonthermal dusty magnetoplasmas with dust-charge fluctuation}
\author{Amar P. Misra}
\homepage{Author to whom any correspondence should be addressed}
\email{apmisra@visva-bharati.ac.in}
\affiliation{Department of Mathematics, Siksha Bhavana, Visva-Bharati University, Santiniketan-731 235, India}
\author{Num P. Acharya}
\email{numacharya@gmail.com}
\affiliation{Central Department of Physics, Tribhuvan University, Kirtipur, Kathmandu 44613, Nepal}
\affiliation{Department of Physics, Mahendra Multiple Campus, Tribhuvan University, Ghorahi 22415, Nepal}
\author{Suresh Basnet}
\email{sbplasma1986@gmail.com}
\affiliation{Asia Pacific Center for Theoretical Physics, Pohang-si, Gyeongsangbuk-do, 37673, South Korea }
\author{Raju Khanal}
\email{raju.khanal@cdp.tu.edu.np}
\affiliation{Central Department of Physics, Tribhuvan University, Kirtipur, Kathmandu 44613, Nepal}
\begin{abstract}
 We reveal a new dispersive dust-lower-hybrid (DLH)-like mode that can couple to modified dust-cyclotron (DC) waves in a dusty magnetoplasma by the influence of a streaming ion beam. Previous studies have overlooked such hybrid modes and the coupling in the study of resonant cyclotron instabilities in dusty magnetoplasmas. Using two-fluid models for positive ion beams and charged dust grains, nonthermal $\kappa$-density distributions for electrons and positive ions, and orbital motion limited (OML) models for dust-charge fluctuations, we derive a general linear dispersion relation for the coupled DLH and DC modes in the limit of when the hydrodynamic time scale is much longer than the dust-charging  time scale. The hybrid mode propagates with a frequency lower than the typical dust-lower-hybrid frequency, $\widetilde{\omega}_{\rm{dl}}\equiv\omega_{\rm{pd}}\Omega_d/\sqrt{\omega^2_{\rm{pd}}+\Omega^2_d}$, where $\omega_{\rm{pd}}$ is the dust-plasma oscillation frequency and $\Omega_d$ is the dust-cyclotron frequency. We obtain the growth rates of instabilities due to Cerenkov and cyclotron interactions and analyze them, taking into account the influences of the static magnetic field, ion-to-electron temperature ratio, electron-to-dust number density ratio, dust-charge fluctuation, and superthermal electrons and ions. We find that the maximum growth rates tend to increase but reach steady states as the wave number increases. The instabilities reported here could be relevant to various plasma environments, including space plasmas (e.g., Earth’s magnetosphere) and laboratory dusty plasma experiments. 
\end{abstract}
\maketitle
\section{Introduction} \label{sec-intro}
 Dusty plasmas are ubiquitous in laboratory \citep{barkan1995laboratory}, as well as in space and astrophysical plasmas such as planetary rings, circumsolar dust grains, interplanetary medium, cometary tails, asteroid zones, Earth's ionosphere and magnetosphere, and in interstellar molecular clouds \citep{mendis1994cosmic,verheest2012waves}. Dust particles can also appear in low-temperature plasmas such as plasma crystals and plasma processing. They become charged through various processes, including photoionization, secondary electron emission, sputtering by energetic ions, etc. \citep{tribeche2010nonlinear}, and their sizes vary from nanometers to micrometers with a dust charge number, $10^2-10^{4}$ as relevant for magnetospheric and ionospheric dusty plasmas \citep{popel2011effect}. Dusty plasmas can support excitation of various waves, including dust-acoustic waves (DAWs) \citep{rao1990dust}, dust-lattice waves (DLWs) \citep{piel2001dynamical}, dust-drift waves (DDWs), and dust-ion-acoustic waves (DIAWs) \citep{shukla1992dust}, and dust-lower hybrid (DLH) waves \cite{amin2006}. 
\par
Space plasmas can exhibit various collective behaviors in the presence of energetic electrons and ions that do not follow Maxwellian distributions (at low energies) but non-Maxwellian ones with high-energy (superthermal) tails. Analysis of different spacecraft measurements confirms the existence of superthermal charged particles \citep{gloeckler2006anisotropic,zouganelis2008measuring,dialynas2009energetic}.  In natural space environments, we describe these charged particles by the Kappa ($\kappa$) or generalized Lorentzian velocity distribution function. The latter well fits the thermal and superthermal components of the observed energy spectra \citep{vasyliunas1969low}. The conventional isotropic three-dimensional form of the generalized Lorentzian or the Kappa distribution function reads \citep{summers1991modified} 
\begin{equation}
\begin{split}
f_{\alpha} (\rm{v})=&\frac{n_{\alpha}}{\left(\kappa_{\alpha}\pi \theta_\alpha^{2}\right)^{3/2}}\frac{\Gamma\left(\kappa_{\alpha}+1\right)}{\Gamma \left(\kappa_{\alpha}-{1}/{2}\right)} \\
&\times \left[1+\frac{\rm{v}^2 }{\kappa_{\alpha}\theta_{\alpha}^{2}}\right]^{-(1+\kappa_{\alpha})},
\end{split} \label{eqn_distribution}
\end{equation}
where $\theta_{\alpha}^{2} = {2T_{\alpha}(\kappa_{\alpha}-{3}/{2})}/{\kappa_{\alpha}m_{\alpha}}$ is the thermal velocity; $n_{\alpha}$, $m_{\alpha}$, and $T_{\alpha}$ (in energy unit) are, respectively, the number density, mass, and temperature of $\alpha$-species superthermal $\kappa$-distributed particles ($\alpha=e$ for electrons and $\alpha=i$ for ions). Also, $\textbf{v}$ = $(\text{v}_{\textrm{x}}, \text{v}_{\textrm{y}}, \text{v}_{\textrm{z}})$ is the particle velocity and $\Gamma(x)$ is the standard Gamma function. 
 \par 
By integrating Eq. \eqref{eqn_distribution} over the velocity space, one can obtain the nonthermal density distributions for $\alpha$-species particles as \citep{tribeche2009effect,tribeche2010nonlinear}
\begin{equation}
n_{\alpha} = n_{\alpha0}\left(1+\frac{Q_{\alpha}\phi}{T_{\alpha}(\kappa_{\alpha}-{3}/{2})}\right)^{-\kappa_{\alpha}+1/2},
\label{eqn-kappa}
\end{equation}
where $n_{\alpha0}$ is the equilibrium number density of electrons ($\alpha=e$) and ions ($\alpha=i$), $Q_\alpha=-e~(e)$ for electrons (ions) with $e$ denoting the elementary charge, and $\phi$ is the electrostatic potential. The spectral index $\kappa_{\alpha}~(>3/2)$ typically measures the steepness of the high-energy tail of the velocity distribution. Lower values of $\kappa_\alpha$ correspond to a steeper tail, indicating more particles with higher energies. However, one can approach the Maxwellian distribution with a less pronounced tail at higher $\kappa_\alpha$-values, i.e., as $\kappa_\alpha\to\infty$. In this context, the  $q$-nonextensive distribution function in Tsallis’ statistics may be somewhat equivalent to the $\kappa_\alpha$-distribution function \eqref{eqn_distribution}, since the latter can be obtained by using the transformation $\kappa_\alpha=q_\alpha/(1-q_\alpha)$ \citep{liu2009dust}.
\par
Studies on dusty plasmas with a mixture of superthermal electrons and ions have gained momentum in recent times due to their significant applications in space and astrophysical environments \citep{hellberg2002generalized,liu2009dust}. Several researchers have focused their attention on such plasmas to study characteristics of DIA and DA waves. To mention a few, Tribeche \textit{et al.} studied arbitrary amplitude DAWs in a dusty plasma with a high-energy-tail electron distribution and the effects of dust-charge fluctuations. They reported that an increase in the electron spectral index results in an increase in the potential amplitude and a decrease in the width of DA solitons. Also, the superthermal electrons can influence the charging of dust grains. The nonlinear theory of electrostatic dust-acoustic solitary waves (DASWs) in unmagnetized dusty plasmas containing nonthermal kappa distributed electrons and thermal ions has been studied. The obtained results indicated that the potential structure is of the compressive type, and the presence of dust temperature and superthermal electrons reduces the dust-acoustic soliton-amplitude \citep{bora2012dust}. In another work, Kaur \textit{et al.} \citep{kaur2017effect} have shown that ion-acoustic solitons in an unmagnetized ion-beam driven plasma with two-temperature kappa-distributed electrons and warm ions can appear as compressive and rarefactive types, whose characteristics get significantly modified by the influence of superthermality of electrons. The influence of kappa-distributed ions on DA solitons can also be found in the work of Habib \textit{et al.} \citep{habib2023dust} in magnetized plasmas containing positively and negatively charged dust particles.   
\par
Typically, the dust-charge fluctuation provides dissipation to the wave motion, leading to linear and nonlinear wave damping or shocks depending on the dust-charging timescale smaller or larger than the inverse of the dust-plasma oscillation frequency. In addition, the charge fluctuation introduces a damped wave eigenmode, whose frequency is proportional to the dust-charging rate. When the dust-charging time scale is much longer than the hydrodynamic time scale, the dust-charge variation can be either slowly varying or almost independent of time, so that its effect cannot contribute to the linear wave modes, but can influence the nonlinear dynamics, which is beyond the scope of the present study. On the other hand, when the dust-charging time scale is much shorter than the hydrodynamic time scale, the dust-charge fluctuation can be assumed to be constant such that the net current flow is zero \cite{verheest2001,avinash1996}. In this case, the dust-charge fluctuation effect is not a dominant damping or instability mechanism.
\par 
The presence of an ion beam in dusty plasmas can influence the dynamics of dust charge fluctuations and, consequently, the wave characteristics. Nejoh \textit{et al.}  \citep{nejoh1999nonlinear} reported that an increase in the ion-beam temperature (density) leads to an increase (decrease) in the dust charge. 
 On the other hand, the resonant interactions of streaming ion beam with magnetoplasmas lead to dust-cyclotron or ion-cyclotron instabilities, which have potential applications in plasma heating and plasma confinement \cite{prakash2014effect,prakash2014ion,sharma1999effect}. However, the streaming ion beam can also couple cyclotron and lower-hybrid-like modes that the previous authors overlooked. Previous studies have largely focused on typical dust-lower-hybrid and dust-cyclotron modes, often neglecting the possible coupling between them, in the presence of ion beam effects. Such coupling can introduce new dispersive hybrid modes that significantly alter the plasma’s collective behavior and energy transport, which is more realistic for natural plasmas where non-Maxwellian effects are important. For instance, Prakash \textit{et al.} \cite{prakash2014ion} studied the characteristics of resonant ion-cyclotron instabilities in an ion-beam driven magnetized dusty plasma without dust-charge fluctuation effects. Also, in another work \cite{prakash2013}, the authors studied lower hybrid waves in electron-beam driven dusty plasmas. However, they have neither identified any lower-hybrid mode nor any coupling between the cyclotron and lower-hybrid modes due to the influence of the ion beam. Although several authors have recently studied the dispersion properties of electrostatic modes, including cyclotron and lower-hybrid modes (See, e.g., Ref. \cite{talukder2025}) and lower-hybrid instabilities (See, e.g., Refs. \cite{munir2024,islam2001}) in dusty plasmas, none of them have considered the coupling between the cyclotron and lower-hybrid modes and ion-beam-driven resonant instabilities.  
 \par 
 In Ref. \cite{islam2001},  Islam \textit{et al.}  considered a Vlasov-kinetic model to mainly study the lower-hybrid wave damping due to Landau resonance and charge fluctuation effects. However, in our work, we introduce new coupling physics by demonstrating the interaction between dust-cyclotron (DC) and dust-lower-hybrid (DLH)-like modes driven by a streaming ion beam, and the associated resonant instabilities (not damping); these features are absent in earlier works. We also include superthermal (kappa-distributed) electrons and ions, take into account the effects of ambient magnetic field and dust-charge fluctuations, and derive growth rates for both Cerenkov and cyclotron resonances, providing broader applicability to space plasmas. Thus, our study extends the previous works by exploring new hybrid modes and their physical implications for plasma heating and cross-field transport mechanisms. 
\section{Basic equations} \label{sec-model}
We consider the propagation of dusty plasma oscillations in a positive ion-beam driven collisionless dusty magnetoplasma composed of nonthermal kappa distributed electrons and ions and negatively charged dust grains with dust-charge fluctuations. We assume the plasma to be immersed in a uniform background magnetic field ${\textbf{B}} = B_{0} \hat{z}$ and ion beam to stream along the magnetic field with a constant velocity ${v}_{\textrm{b}0}$ in the background plasma. At equilibrium, the charge quasineutrality condition reads
\begin{equation}
 en_{\mathrm{e0}} - en_{\textrm{i}0}-en_{\textrm{b}0}-q_{d0}n_{\mathrm{d0}} = 0,
\end{equation}
where $n_{\alpha0}$ is the equilibrium number density of $\alpha$-species particle ($\alpha$ = $i$, $e$, $b$, and $d$ are for positive ions, electrons, positive ion beam, and charged dust grains) and $q_{\textrm{d}0}$ is the equilibrium dust charge.  
The basic fluid equations for negatively charged dust grains ($\alpha=d$) and positive ion beam ($\alpha=b$) are
\begin{equation}
\frac{\partial n_{\alpha}}{\partial t} + \nabla\cdot\left(n_{\alpha}{\textbf{v}_{\alpha}}\right) = 0,
\label{dust_continuity}
\end{equation}
\begin{equation}
\begin{split}
m_{\alpha}n_{\mathrm{\alpha}} &\left(\frac{\partial {\textbf{v}_{\alpha}}}{\partial t} + \left({\textbf{v}_{\alpha}}\cdot\nabla \right){\textbf{v}_{\alpha}}\right)  = -q_{\alpha}n_{\alpha}\nabla\phi \\
&+ q_{\alpha} n_{\alpha}\left({\textbf{v}_{\alpha}}\times {\textbf{B}}\right) 
- \nabla P_{\alpha},
\label{eqn_momentum}
\end{split}
\end{equation}
where $q_d=-Z_de$ with $Z_d$ denoting the dust-charge number and $q_b=e$.
\par
The above system of equations are closed by the superthermal distributions of electrons and ions [See Eq. \eqref{eqn-kappa}],  Poisson's equation, and the pressure law, given by,
\begin{equation}
{\varepsilon_0}\nabla^{2}\phi = \left(en_{\mathrm{e}}-en_{\mathrm{i}}-en_{\textrm{b}}-q_{\mathrm{d}}n_{\mathrm{d}}\right),
\label{eqn_Poisson}
\end{equation}
\begin{equation}
P_\alpha=P_{\alpha0}\left(\frac{n_\alpha}{n_{\alpha0}}\right)^\gamma
\end{equation}
where $P_\alpha$ is the thermodynamic pressure with $P_{\alpha}=P_{\alpha0}$ at $n_\alpha=n_{\alpha0}$,   $\varepsilon_0$ is the permittivity of free space, and $\gamma$ is the adiabatic index.
\par
Spherical dust grains with constant radius $r_{\textrm{d}}$ collect electrons, positive ions, and positive ion beams onto their surfaces and become charged. In space plasmas, the dust particles are negatively charged. This is due to the larger flow of plasma electrons from the background plasma onto the dust-grain surfaces \citep{shukla2001survey}. The dust charging equation is given by \citep{nejoh1999nonlinear}
\begin{equation}
\frac{dq_{\textrm{d}}}{dt} = I_{\textrm{e}}+ I_{\textrm{i}} + I_{\textrm{b}},
 \label{eqn_dustcharging}
\end{equation}
where the several currents that flow onto the dust grain surface are due to $\kappa$-distributed electron ($I_{\textrm{e}}$), $\kappa$-distributed positive ion ($I_{\textrm{i}}$), and positive ion beam ($I_{\textrm{b}}$) currents. According to the orbital motion limited (OML) theory, the superthermal electron and ion currents entering the dust-grain surface are given by \citep{tribeche2009effect,tribeche2010nonlinear}
\begin{equation}
\begin{split}
I_{\textrm{e}} =& -\gamma_{1}\pi r_{\textrm{d}}^{2}en_{\textrm{e}0}\sqrt{\frac{8T_{\textrm{e}}}{\pi m_{\textrm{e}}}} \\
&\times\left[1-\frac{1}{k_{\textrm{e}}-{3}/{2}}\left(\frac{eq_{\textrm{d}}}{r_{\textrm{d}}T_{\textrm{e}}}+\frac{e\phi}{T_{\textrm{e}}}\right)\right]^{1-\kappa_{\textrm{e}}},
\end{split}
\end{equation}
\begin{equation}
\begin{aligned}
I_{\mathrm{i}} = \gamma_{2}\pi r^{2}_{\mathrm{d}}en_{\mathrm{i0}}\sqrt{\frac{8T_{\mathrm{i}}}{\pi m_{\mathrm{i}}}}\biggl[\left(1+\frac{e\phi}{T_{\textrm{i}}\left(\kappa_{\textrm{i}}-{3}/{2}\right)}\right)^{1-\kappa_{\textrm{i}}}\\ -\frac{\kappa_{\textrm{i}}-1}{\kappa_{\textrm{i}}-{3}/{2}}\frac{eq_{\textrm{d}}}{r_{\textrm{d}}T_{\textrm{i}}}\left(1+\frac{e\phi}{T_{\textrm{i}}\left(\kappa_{\textrm{i}}-{3}/{2}\right)}\right)^{-\kappa_{\textrm{i}}}\biggr],
\label{eqn_ion_current}
\end{aligned}
\end{equation}
where the expressions for $\gamma_1$ and $\gamma_2$ are 
\begin{equation}
\begin{split}
&\gamma_{1} = \frac{\Gamma(\kappa_{\textrm{e}}+1)}{\Gamma (\kappa_{\textrm{e}}-{1}/{2})}\frac{(\kappa_{\textrm{e}}-{3}/{2})^{{1}/{2}}}{\kappa_{\textrm{e}}(\kappa_{\textrm{e}}-1)},\\
&\gamma_{2} = \frac{\Gamma(\kappa_{\textrm{i}}+1)}{\Gamma (\kappa_{\textrm{i}}-{1}/{2})}\frac{(\kappa_{\textrm{i}}-{3}/{2})^{{1}/{2}}}{\kappa_{\textrm{i}}(\kappa_{\textrm{i}}-1)}.
\end{split}
\end{equation}
Also, if the beam thermal velocity $\text{v}_{\textrm{tb}}$ is much smaller than the beam streaming velocity $\text{v}_{\textrm{b}0}$, then the beam current becomes independent of the beam temperature $(T_b)$ and its expression can be approximated as \citep{shukla2001survey}
\begin{equation}
I_{\mathrm{b}} = \pi r^{2}_{\mathrm{d}}e n_{\mathrm{b}}\text{v}_{\textrm{b}0}\left[1- \frac{2eq_{\textrm{d}}}{r_{\textrm{d}}m_{\textrm{b}}\text{v}_{\textrm{b}0}^{2}}\right].
\label{eqn_beam_current}
\end{equation}
\section{Dispersion relation} \label{sec-disp}
We assume that the wave amplitudes of dusty plasma oscillations are small such that terms involving second or higher order perturbations are negligible. To obtain the linear dispersion relation, we perturb the physical quantities into its equilibrium (with a zero value or a suffix $0$) and perturbation parts (with a suffix $1$), i.e., $\left(n_d, \rm{v}_d, \rm{v}_b, \phi, q_d\right)=\left(n_{\textrm{d}0}, 0,  \text{v}_{\textrm{b}0}, 0, q_{\textrm{d}0}\right)$, and assume the perturbed quantities to vary as plane waves of constant amplitudes in the form $\exp[i({\bf k}\cdot {\bf r}-\omega t)]$ with the wave vector ${\bf k}$ and wave frequency $\omega$. 
Here, ${\bf{r}} = (x,\text{y},\text{z})$ and   $k^{2} = k_{\textrm{x}}^{2}+k_{\textrm{y}}^{2}+k_{\textrm{z}}^{2} = k_{\perp}^{2}+k_{\parallel}^{2}$ with $k_{\perp}^{2}=k_{\textrm{x}}^{2}+k_{\textrm{y}}^{2}$ and $k_{\parallel}^{2}=k_{\textrm{z}}^{2}$.
Thus, from  Eqs. (\ref{dust_continuity})-(\ref{eqn_Poisson}) we obtain the following linear equations for the perturbed quantities.
\begin{equation}
n_{\textrm{d}1} = \frac{k_{\textrm{x}}}{\omega}n_{\textrm{d}0}\text{v}_{\textrm{dx}1}+\frac{k_{\textrm{y}}}{\omega}n_{\textrm{d}0}\text{v}_{\textrm{dy}1}+\frac{k_{\textrm{z}}}{\omega}n_{\textrm{d}0}\text{v}_{\textrm{dz}1},
\label{eqn_dust_linear}
\end{equation}
\begin{equation}
\text{v}_{\textrm{dx}1}=\frac{q_{\textrm{d}0}k_{\textrm{x}}}{m_{\textrm{d}}\omega}\phi_{1}+i\frac{q_{\textrm{d}0}B}{m_{\textrm{d}}\omega}\text{v}_{\textrm{dy}1}+\text{v}_{\textrm{td}}^{2}\frac{k_{\textrm{x}}}{\omega}\frac{n_{\textrm{d}1}}{n_{\textrm{d}0}},
\label{eqn_velocity_xcomp}
\end{equation}
\begin{equation}
\text{v}_{\textrm{dy}1}=\frac{q_{\textrm{d}0}k_{\textrm{y}}}{m_{\textrm{d}}\omega}\phi_{1}-i\frac{q_{\textrm{d}0}B}{m_{\textrm{d}}\omega}\text{v}_{\textrm{dx}1}+\text{v}_{\textrm{td}}^{2}\frac{k_{\textrm{y}}}{\omega}\frac{n_{\textrm{d}1}}{n_{\textrm{d}0}},
\label{eqn_velocity_ycomp}
\end{equation}
\begin{equation}
\text{v}_{\textrm{dz}1}=\frac{q_{\textrm{d}0}k_{\textrm{z}}}{m_{\textrm{d}}\omega}\phi_{1}+\text{v}_{\textrm{td}}^{2}\frac{k_{\textrm{z}}}{\omega}\frac{n_{\textrm{d}1}}{n_{\textrm{d}0}},
\label{eqn_velocity_zcomp}
\end{equation}
\begin{equation}
n_{\textrm{b}1} = \frac{k_{\textrm{x}}}{\omega-k_{\textrm{z}}\text{v}_{\textrm{b}0}}n_{\textrm{b}0}\text{v}_{\textrm{bx}1}+\frac{k_{\textrm{y}}n_{\textrm{b}0}}{\omega-k_{\textrm{z}}\text{v}_{\textrm{b}0}}\text{v}_{\textrm{by}1}+\frac{k_{\textrm{z}}n_{\textrm{b}0}}{\omega-k_{\textrm{z}}\text{v}_{\textrm{b}0}}\text{v}_{\textrm{bz}1},
\label{eqn_beam_linear}
\end{equation}
\begin{equation}
\text{v}_{\textrm{bx}1}= \frac{e}{m_{\textrm{b}}}\frac{k_{\textrm{x}}}{\omega-k_\textrm{z}\text{v}_{\textrm{b}0}}\phi_{1}+i\frac{eB}{m_{\textrm{b}}}\frac{\text{v}_{\textrm{by}1}}{\omega-k_{\textrm{z}}\text{v}_{\textrm{b}0}}+\text{v}_{\textrm{tb}}^{2}\frac{k_{\textrm{x}}}{\omega-k_{\textrm{z}}\text{v}_{\textrm{b}0}}\frac{n_{\textrm{b}1}}{n_{\textrm{b}0}},
\label{eqn_velocity_xcomp_beam}
\end{equation}
\begin{equation}
\text{v}_{\textrm{by}1}= \frac{e}{m_{\textrm{b}}}\frac{k_{\textrm{y}}}{\omega-k_{\textrm{z}}\text{v}_{\textrm{b}0}}\phi_{1}-i\frac{eB}{m_{\textrm{b}}}\frac{\text{v}_{\textrm{bx}1}}{\omega-k_{\textrm{z}}\text{v}_{\textrm{b}0}}+\text{v}_{\textrm{tb}}^{2}\frac{k_{\textrm{y}}}{\omega-k_{\textrm{z}}\text{v}_{\textrm{b}0}}\frac{n_{\textrm{b}1}}{n_{\textrm{b}0}},
\label{eqn_velocity_ycomp_beam}
\end{equation}
\begin{equation}
\text{v}_{\textrm{bz}1}= \frac{e}{m_{\textrm{b}}}\frac{k_{\textrm{z}}}{\omega-k_{\textrm{z}}\text{v}_{\textrm{b}0}}\phi_{1}+\text{v}_{\textrm{tb}}^{2}\frac{k_{\textrm{z}}}{\omega-k_{\textrm{z}}\text{v}_{\textrm{b}0}}\frac{n_{\textrm{b}1}}{n_{\textrm{b}0}},
\label{eqn_velocity_zcomp_beam}
\end{equation}
\begin{equation}
    -k^{2}\varepsilon_{0}\phi_{1} = en_{\textrm{e}1}-en_{\textrm{i}1}-en_{\textrm{b}1}-q_{\textrm{d}0}n_{\textrm{d}1}-n_{\textrm{d}0}q_{\textrm{d}1},
    \label{perturbed_poisson}
\end{equation}
where $\text{v}_{\mathrm{t}\alpha} = \sqrt{\gamma T_{\alpha}/m_{\alpha}}$ is the thermal velocity of $\alpha$-species particle with $\gamma = 5/3$ for three dimensional wave propagation.
\par
From Eqs. (\ref{eqn_dust_linear}) - (\ref{eqn_velocity_zcomp_beam}), solving for the perturbed densities, we have
\begin{equation} n_{\textrm{d}1}=\frac{n_{\textrm{d}0}q_{\textrm{d}0}}{m_{\textrm{d}}}\frac{\frac{k_{\perp}^{2}}{\omega^{2}-\Omega_d^{2}}+\frac{k_{\parallel}^{2}}{\omega^{2}}}{1-\text{v}_{\textrm{td}}^{2}\left(\frac{k_{\perp}^{2}}{\omega^{2}-\Omega_d^{2}}+\frac{k_{\parallel}^{2}}{\omega_d^{2}}\right)}\phi_{1},
\label{eqn_dust_density_perturbed}
\end{equation}
\begin{equation} n_{\textrm{b}1}=\frac{en_{\textrm{b}0}}{m_{\textrm{b}}}\frac{\frac{k_{\perp}^{2}}{\widetilde{\omega}^{2}-\Omega_{\textrm{b}}^{2}}+\frac{k_{\parallel}^{2}}{\widetilde{\omega}^{2}}}{1-\text{v}_{\textrm{td}}^{2}\left(\frac{k_{\perp}^{2}}{\widetilde{\omega}^{2}-\Omega_{\textrm{b}}^{2}}+\frac{k_{\parallel}^{2}}{\widetilde{\omega}^{2}}\right)}\phi_{1},
\label{eqn_beam_density_perturbed}
\end{equation}
\begin{equation}
n_{\textrm{i}1} = -n_{\textrm{i}0} \frac{\kappa_{\textrm{i}}-{1}/{2}}{\kappa_{\textrm{i}}-{3}/{2}}  \frac{e\phi_{1}}{T_{\mathrm{i}}},
\label{perturbed_ion_density}
\end{equation}
\begin{equation}
n_{\textrm{e}1} = n_{\textrm{e}0} \frac{\kappa_{\textrm{e}}-{1}/{2}}{\kappa_{\textrm{e}}-{3}/{2}}  \frac{e\phi_{1}}{T_{\mathrm{e}}},
\label{perturbed_electron_density}
\end{equation}
where $\Omega_d ={|q_{\textrm{d}0}|B_{0}}/{m_{\textrm{d}}}$ and $\Omega_\textrm{b} = {eB_{0}}/{m_{\textrm{b}}}$, respectively, denote the dust- and beam-cyclotron frequencies and $\widetilde{\omega} = \omega-k_{\parallel}\text{v}_{\textrm{b}0}$ is the frequency downshift due to streaming ion beam parallel to the magnetic field.  From Eqs. \eqref{eqn_dust_density_perturbed} and \eqref{eqn_beam_density_perturbed}, we note that both the dust-cyclotron (at $\omega=\Omega_d$) and beam-cyclotron resonances (at $\omega=k_\parallel \rm{v}_{\rm{b0}}\pm \Omega_b$) are associated with the perpendicular wave propagation of dust and beam density perturbations.
 However, as expected, there is no resonance for longitudinal density perturbations along the magnetic field and thus the dust and beam density perturbations are acoustic in character for parallel wave propagation. 
\par
The equilibrium state of dust grains gives the equilibrium dust charge ($q_{\textrm{d}0}$). At equilibrium, the net current flowing onto the dust grain surface is zero, i.e., 
\begin{equation}
I_{\textrm{e}0}+I_{\textrm{i}0}+I_{\textrm{b}0}=0,
\label{eqn_equilibrium_current}
\end{equation}
where the electron, ion, and beam currents are
\begin{equation}
I_{\textrm{e}0}=-\gamma_{1}\pi r_{\textrm{d}}^{2}en_{\textrm{e}0}\sqrt{\frac{8T_{\textrm{e}}}{\pi m_{\textrm{e}}}}\left(1-\frac{eq_{\textrm{d}0}}{(\kappa_{\textrm{e}}-{3}/{2})r_{\textrm{d}T_{\textrm{e}}}}\right)^{1-\kappa_{\textrm{e}}},
\label{eqm_electron_current}
\end{equation}
\begin{equation}
I_{\textrm{i}0}=\gamma_{2}\pi r_{\textrm{d}}^{2}en_{\textrm{i}0}\sqrt{\frac{8T_{\textrm{i}}}{\pi m_{\textrm{i}}}}\left(1-\frac{(\kappa_{\textrm{i}}-1)eq_{\textrm{d}0}}{(\kappa_{\textrm{i}}-{3}/{2})r_{\textrm{d}T_{\textrm{i}}}}\right),
\label{eqm_ion_current}
\end{equation}
\begin{equation}
I_{\textrm{b}0}= \pi r_{\textrm{d}}^{2}en_{\textrm{b}0}\text{v}_{\textrm{b}0}\left(1-\frac{2eq_{\textrm{d}0}}{r_{\textrm{d}}m_{\textrm{b}}\text{v}_{\textrm{b}0}^{2}}\right).
\label{eqm_beam_current}
\end{equation}
However, due to collisions of dust grains with electrons, ions, and beam, the dust charge gets perturbed.  So, the linear perturbation equation for dust charge fluctuations is given by
\begin{equation}
\frac{dq_{\textrm{d}1}}{dt} = I_{\textrm{i}1} + I_{\textrm{b}1}+I_{\textrm{e}1}, 
\label{eqn_fluctuation_dustcharging}
\end{equation}
where the fluctuating currents are
\begin{equation}
\begin{split}
I_{\textrm{e}1} =& I_{\textrm{e}0} \frac{\kappa_{\textrm{e}}-1}{\kappa_{\textrm{e}}-{3}/{2}}\left[{1-\frac{eq_{\textrm{d}0}}{\left(\kappa_{\textrm{e}}-{3}/{2}\right)r_{\textrm{d}}T_{\textrm{e}}}}\right]^{-1}\\
&\times\left(\frac{eq_{\textrm{d}1}}{r_{\textrm{d}}T_{\textrm{e}}}+\frac{e\phi_{1}}{T_{\textrm{e}}}\right),
\label{eqn_first_flu_electron_current}
\end{split}
\end{equation}
\begin{equation}
\begin{split}
I_{\textrm{i}1} = &I_{\textrm{i}0} \frac{\kappa_{\textrm{i}}-1}{\kappa_{\textrm{i}}-{3}/{2}}\left[{1-\frac{\left(\kappa_{\textrm{i}}-1\right)eq_{\textrm{d}0}}{\left(\kappa_{\textrm{i}}-{3}/{2}\right)r_{\textrm{d}}T_{\textrm{i}}}}\right]^{-1}\\
&\times\biggl[-\frac{eq_{\textrm{d}1}}{r_{\textrm{d}}T_{\textrm{e}}}+  \left(\frac{\kappa_{\textrm{i}}eq_{\textrm{d}0}}{(\kappa_{\textrm{i}}-{3}/{2})r_{\textrm{d}}T_{\textrm{i}}}-1\right)\frac{e\phi_{1}}{T_{\textrm{e}}}\biggr],
\label{eqn_first_flu_ion_current}
\end{split}
\end{equation}
\begin{equation}
I_{\textrm{b}1} = I_{\textrm{b}0} \left[\frac{n_{\textrm{b}1}}{n_{\textrm{b}0}}-\frac{2eq_{\textrm{d}1}}{r_{\textrm{d}}m_{\textrm{b}}\text{v}_{\textrm{b}0}^{2}}\left({1-\frac{2eq_{\textrm{d}0}}{r_{\textrm{d}}m_{\textrm{b}}\text{v}_{\textrm{b}0}^2}}\right)^{-1}\right].
\label{eqn_first_flu_beam_current}
\end{equation}
Substituting Eqs. (\ref{eqn_beam_density_perturbed}), (\ref{eqn_first_flu_electron_current})--(\ref{eqn_first_flu_beam_current}) into Eq. (\ref{eqn_fluctuation_dustcharging}), we obtain the following expression for the perturbed dust charge ($q_{\textrm{d}1}$).
\begin{equation}
\begin{aligned}
q_{\textrm{d}1} =\frac{i}{\omega+i\eta}\biggl[I_{\textrm{i}0}\beta_{i}\left(\frac{\kappa_{\textrm{i}}}{\kappa_{\textrm{i}}-\frac{3}{2}}\frac{eq_{\textrm{d}0}}{r_{\textrm{d}}T_{\textrm{i}}}-1\right)\frac{e\phi_{1}}{T_{\textrm{i}}}+\\I_{\textrm{e}0}\beta_{e}\frac{e\phi_{1}}{T_{\textrm{e}}}+I_{\textrm{b}0}\frac{n_{\textrm{b}1}}{n_{\textrm{b}0}}\biggr],
\label{eqn_perturbed_dustcharge}
\end{aligned}
\end{equation}
where 
\begin{equation}
\begin{split}
&\beta_{i} = \frac{\kappa_{\textrm{i}}-1}{(\kappa_{\textrm{i}}-{3}/{2})}\left[{1-\frac{(\kappa_{\textrm{i}}-1)eq_{\textrm{d}0}}{(\kappa_{\textrm{i}}-{3}/{2})r_{\textrm{d}}T_{\textrm{i}}}}\right]^{-1},\\
& \beta_{e} = \frac{\kappa_{\textrm{e}}-1}{\kappa_{\textrm{e}-{3}/{2}}}\left[{1-\frac{eq_{\textrm{d}0}}{(\kappa_{\textrm{e}}-{3}/{2})r_{\textrm{d}}T_{\textrm{e}}}}\right]^{-1},
\end{split}
\end{equation}
and the dust charging rate ($\eta$) is given by
\begin{equation}
\eta = \frac{e}{r_{\textrm{d}}T_{\textrm{e}}}\left(I_{\textrm{i}0}\frac{\beta_{i}}{\sigma_{\textrm{i}}}-I_{\textrm{e}0}\beta_e+I_{\textrm{b}0}\beta_{b}\right).
\label{eqn_charging_rate}
\end{equation}
Here, $\sigma_{\textrm{i}} = {T_{\textrm{i}}}/{T_{\textrm{e}}}$ and $\beta_b$ is given by
\begin{equation}
  \beta_{b} = \frac{2T_\textrm{e}}{m_{\textrm{b}}\text{v}_{\textrm{b}0}^{2}}\left({1-\frac{2eq_{\textrm{d}0}}{r_{\textrm{d}}m_{\textrm{b}}\text{v}_{\textrm{b}0}^{2}}}\right)^{-1}.
  \end{equation}
\par
Finally, substituting Eqs. (\ref{eqn_dust_density_perturbed})--(\ref{perturbed_electron_density}) and (\ref{eqn_perturbed_dustcharge}) into Eq. (\ref{perturbed_poisson}), we obtain the following general dispersion relation for coupled DC and DLH-like modes in superthermal magnetized dusty plasmas.
\begin{equation} \label{eqn_linear_dispersion}
\begin{split}
1+\frac{1}{\lambda_{\textrm{D}}^{2}k^{2}} & -\frac{i}{\omega+i\eta}\frac{1}{k^{2}}\left[\frac{\alpha_{\textrm{e}}\beta_{e}}{\lambda_{\textrm{De}}^{2}}+\frac{\alpha_{\textrm{i}}\beta_{i}}{\lambda_{\textrm{Di}}^{2}}\left(\frac{k_{\textrm{i}}}{k_{\textrm{i}}-3/2}\frac{eq_{\textrm{d}0}}{r_{\textrm{d}}T_{\textrm{i}}}-1\right)\right]\\
- & \frac{\omega_{\textrm{pb}}^{2}}{k^{2}}\chi_b\left(1+\frac{i\alpha_b}{\omega+i\eta}\right)-\frac{\omega_{\textrm{pd}}^{2}}{k^{2}}\chi_d=0,
\end{split}
\end{equation}
where $\alpha_e,~\alpha_i,~\alpha_b,~\chi_b$, and $\chi_d$ are given by
\begin{equation*}
\begin{aligned}
\alpha_{\textrm{e}} = \frac{I_{\textrm{e}0}}{e}\frac{n_{\textrm{d0}}}{n_{\textrm{e}0}},~
\alpha_{\textrm{i}} = \frac{I_{\textrm{i}0}}{e}\frac{n_{\textrm{d0}}}{n_{\textrm{i}0}},~
\alpha_{\textrm{b}} = \frac{I_{\textrm{b}0}}{e}\frac{n_{\textrm{d0}}}{n_{\textrm{b}0}},
\end{aligned}
\end{equation*}
\begin{equation}
\chi_b=\left({\frac{k_{\perp}^{2}}{\widetilde{\omega}^{2}-\Omega_{\textrm{b}}^{2}}+\frac{k_{\parallel}^{2}}{\widetilde{\omega}^{2}}}\right)\Bigg/\left[{1-\text{v}_{\textrm{tb}}^{2}\left(\frac{k_{\perp}^{2}}{\widetilde{\omega}^{2}-\Omega_{\textrm{b}}^{2}}+\frac{k_{\parallel}^{2}}{\widetilde{\omega}^{2}}\right)}\right],
\end{equation}
\begin{equation}
\chi_d=\left({\frac{k_{\perp}^{2}}{\omega^{2}-\Omega_d^{2}}+\frac{k_{\parallel}^{2}}{\omega^{2}}}\right)\Bigg/\left[{1-\text{v}_{\textrm{td}}^{2}\left(\frac{k_{\perp}^{2}}{\omega^{2}-\Omega_d^{2}}+\frac{k_{\parallel}^{2}}{\omega^{2}}\right)}\right].
\end{equation}
Also, $\lambda_D$ is the effective Debye length, given by,
\begin{equation}
\frac{1}{\lambda_{\textrm{D}}^{2}} = \frac{1}{\lambda_{\textrm{Di}}^{2}}\frac{\kappa_{\textrm{i}}-{1}/{2}}{\kappa_{\textrm{i}}-{3}/{2}}+ \frac{1}{\lambda_{\textrm{De}}^{2}}\frac{\kappa_{\textrm{e}}-1/2}{\kappa_{\textrm{e}}-3/2},
\end{equation}
where $\lambda_{\textrm{Dj}} = \left(\gamma\varepsilon_{0}T_{\textrm{j}}/n_{\textrm{j}0}e^{2}\right)^{1/2}$ is the Debye length of $j$-th species particle with $j=e~(i)$ for electrons (ions).
 \par 
Typically, for cyclotron modes, $\Omega_\alpha\gg k\text{v}_{\textrm{t}\alpha}$ $(\alpha=d,~b)$. Also, we have assumed that $\rm{v}_{\rm{b0}}\gg \rm{v}_{\rm{tb}}$. Thus, in Eq. \eqref{eqn_linear_dispersion}, we can neglect the thermal contributions from charged dusts and the ion beam. In this case, Eq. (\ref{eqn_linear_dispersion}) reduces to
\begin{equation}
\begin{split}
\left(\omega^{2}-\omega_{\textrm{dc}}^{2}\right)&\left(\omega^{2}-\omega_{\textrm{dl}}^{2}\right) = \omega^{2}(\omega^{2}-\Omega_d^{2})\frac{\omega_{\textrm{pb}}^{2}\lambda_{\textrm{D}}^{2}}{1+k^{2}\lambda_{\textrm{D}}^{2}}\\
&\times \left(1+\frac{i\alpha_b}{\omega+i\eta}\right) \left(\frac{k_\perp^{2}}{{\widetilde{\omega}^{2}}-\Omega_{\textrm{b}}^{2}}+\frac{k_\parallel^{2}}{{\widetilde{\omega}^{2}}}\right),
\label{eq-DR1}
\end{split}
\end{equation}
where $\omega_{\textrm{dc}}$ and $\omega_{\textrm{dl}}$ are, respectively, the frequencies associated with the dust-cyclotron and dust-lower-hybrid-like modes, given by,
\begin{equation}
\omega_{\textrm{dc}}^{2} = \Omega_d^{2} + \frac{k^{2}C_{\textrm{D}}^{2}}{1+k^{2}\lambda_{\textrm{D}}^{2}}+\frac{i{\widetilde{\alpha}_{\textrm{c}}}}{\omega+i\eta}\frac{\omega^{2}-\Omega_d^{2}}{1+k^{2}\lambda_{\textrm{D}}^{2}},
\label{eq-dc1}
\end{equation}
\begin{equation} \label{eq-dl1}
\frac{1}{\omega_{\textrm{dl}}^{2}}=\frac1{\omega_{\textrm{da}}^{2}}+\frac{1}{\Omega_d^{2}}
>\frac{1}{\widetilde{\omega}_{\textrm{dl}}^{2}}\equiv \frac1{\omega_{\textrm{pd}}^{2}}+\frac{1}{\Omega_d^{2}},
\end{equation}
where $\omega_{\textrm{da}}=kC_D/\sqrt{1+k^2\lambda_{\textrm{D}}^{2}}~(<\omega_{\textrm{pd}})$ is the frequency of dust-acoustic mode 
in a dusty plasma with $C_D=\omega_{\textrm{pd}}\lambda_D$ denoting the dust-acoustic speed. Also, we have the following expression for $\widetilde{\alpha}_c$.
\begin{equation} \label{eq-alphac}
\widetilde{\alpha}_c=\alpha_e\beta_e\frac{\lambda_D^2}{\lambda_{De}^2}+\alpha_i\beta_i\frac{\lambda_D^2}{\lambda_{Di}^2} \left(\frac{\kappa_i}{\kappa_i-3/2}\frac{e q_{d0}}{r_dT_i}-1\right).
\end{equation}
  In the absence of the dust-charge fluctuation effect, Eq. \eqref{eq-DR1} agrees with the previous work \citep{prakash2014ion} after one takes the limits $\kappa_{\textrm{i}}, \kappa_{\textrm{e}} \to \infty$ for Maxwellian electrons and ions, and replaces the dust-plasma frequency $\omega_{\textrm{pd}}$ by the ion-plasma frequency $\omega_{\textrm{pi}}$ and dust-cyclotron frequency $(\Omega_d)$ by the ion-cyclotron frequency $(\omega_{\textrm{ci}})$. 
From Eq. \eqref{eq-DR1}, it is interesting to note that not only does DC mode exist, but there also exists another low-frequency DLH-like mode. This mode is dispersive and propagates with a frequency lower than the typical DLH frequency $\widetilde{\omega}_{\rm{dl}}$. Moreover, the DC and DLH-like modes are coupled by the influence of the positive ion beam in a dusty magnetoplasma, unless $\omega=\Omega_d$ (the case of constant oscillation). Previous works have overlooked the existence of such lower-hybrid modes and their coupling with the cyclotron modes in the study of ion-beam-driven cyclotron instabilities \cite{prakash2014effect,prakash2014ion}. From Eq. \eqref{eq-DR1}, we also note that in the absence of the ion beam, there will be no resonance. In this case, we have only the decoupled modes: (i) $\omega=\omega_{\rm{dc}}$, the DC mode modified by the dust-charge fluctuation effect, and (ii) the DLH-like mode $\omega=\omega_{\rm{dl}}$. Another noteworthy point is that for the coupled modes, resonances can occur at frequencies $\omega=\omega_0\equiv k_\parallel \rm{v_{b0}}$ and $\omega=\omega_{\pm}=k_\parallel \rm{v_{b0}}\pm \Omega_b$. The former ($\omega_0$) is associated with the streaming ion beam. The latter ($\omega_\pm$) is caused by the combined influence of the streaming ion beam and the beam cyclotron motion. However, between the two frequencies $\omega_{\pm}$, the wave will more easily acquire the lower frequency. As a result, the resonance at $\omega=\omega_{-}$ will dominate over that at $\omega=\omega_{+}$. Therefore, we neglect the resonant instability associated with $\omega=\omega_{+}$. 
\par 
We are interested in the instabilities of the coupled DC and DLH modes due to the resonances at $\omega=\omega_0$ and $\omega=\omega_-$. We consider the dust-charging time scale to be much shorter than the hydrodynamic time scale, i.e., $\eta\gg\omega_{\rm{pd}}$ for which the dust-charge fluctuation can become constant such that the net current flow is zero, and the charge fluctuation effect is not a dominant instability or damping mechanism. In the case of $\eta\gg\omega_{\rm{pd}}$, the dispersion equation \eqref{eq-DR1} and Eq. \eqref{eq-dc1} further reduce  to 
\begin{equation}
\begin{split}
\left(\omega^{2}-\omega_{\textrm{dc}}^{2}\right)&\left(\omega^{2}-\omega_{\textrm{dl}}^{2}\right)\approx \omega^{2}(\omega^{2}-\Omega_d^{2})\frac{\omega_{\textrm{pb}}^{2}\lambda_{\textrm{D}}^{2}}{1+k^{2}\lambda_{\textrm{D}}^{2}}\\
&\times \left(1+\frac{\alpha_b}{\eta}\right) \left(\frac{k_\perp^{2}}{{\widetilde{\omega}^{2}}-\Omega_{\textrm{b}}^{2}}+\frac{k_\parallel^{2}}{{\widetilde{\omega}^{2}}}\right),
\label{eq-DR2}
\end{split}
\end{equation}
where
\begin{equation}
\omega_{\textrm{dc}}^{2} \approx \widetilde{\omega}_{\textrm{dc}}^{2}+\frac{{\widetilde{\alpha}_{\textrm{c}}}}{\eta}\frac{k^{2}C_{\textrm{D}}^{2}}{\left(1+k^{2}\lambda_{\textrm{D}}^{2}\right)^2},
\label{eq-dc2}
\end{equation}
with \begin{equation}
\widetilde{\omega}_{\textrm{dc}}^{2}=\Omega_d^{2} + \frac{k^{2}C_{\textrm{D}}^{2}}{1+k^{2}\lambda_{\textrm{D}}^{2}}
\end{equation}
denoting the squared dust-cyclotron wave frequency in the absence of dust-charge fluctuation.
\section{Growth rate of instability} \label{sec-instability}
In Cerenkov interaction, we consider the resonance at $\omega=\omega_0\equiv k_\parallel \rm{v_{b0}}$. So, retaining the term proportional to $k^2_\parallel$ $(\gg k_\perp^2)$ in Eq. \eqref{eq-DR2}, and assuming $\omega=k_\parallel \rm{v_{b0}}+\delta$, where $\delta~(\ll\omega_0)$ is the small frequency mismatch associated with the instability, we obtain the following growth rates of instabilities $(\Im\delta)$ corresponding to DC and DLH modes.
\begin{equation}\label{eq-growth-dc1}
\begin{split}
\gamma_{\rm{dc}} \approx &\left[\frac{\omega_{\rm{dc}}^{2}-\Omega_{\rm{d}}^{2}}{\omega_{\rm{dc}}^{2}-\omega_{\rm{dl}}^{2}}\left(1+\frac{\alpha_b}{\eta}\right)\frac{\omega_{\rm{pb}}^{2}k^2_\parallel\lambda^2_{\rm{D}}}{1+k^2_\parallel\lambda^2_{\rm{D}}} \right]^{1/2} \\
&\times \left(1-\frac{k^2_\parallel \rm{v^2_{{b0}}}}{\omega_{\rm{dc}}^{2}}\right)^{-1/2},
\end{split}
\end{equation}
\begin{equation}\label{eq-growth-dl1}
\begin{split}
\gamma_{\rm{dl}} \approx &\left[\frac{\omega_{\rm{dl}}^{2}-\Omega_{\rm{d}}^{2}}{\omega_{\rm{dl}}^{2}-\omega_{\rm{dc}}^{2}}\left(1+\frac{\alpha_b}{\eta}\right)\frac{\omega_{\rm{pb}}^{2}k^2_\parallel\lambda^2_{\rm{D}}}{1+k^2_\parallel\lambda^2_{\rm{D}}} \right]^{1/2} \\
&\times \left(1-\frac{k^2_\parallel \rm{v^2_{{b0}}}}{\omega_{\rm{dl}}^{2}}\right)^{-1/2},
\end{split}
\end{equation} 
From Eqs. \eqref{eq-growth-dc1} and \eqref{eq-growth-dl1}, we note that the growth rates can achieve maximum values when both the DC and DLH frequencies approach the frequency $\omega_0$, i.e., 
$\omega_{\rm{dc}}\approx k_\parallel \rm{v_{b0}}$ and $\omega_{\rm{dl}}\approx k_\parallel \rm{v_{b0}}$. So, to obtain the maximum growth rates for the DC and DLH modes, we must consider the resonances at $\omega=k_\parallel \rm{v_{b0}}$ and $\omega= \omega_{\rm{dc}}$ for the DC instability and $\omega=k_\parallel \rm{v_{b0}}$ and $\omega= \omega_{\rm{dl}}$ for the DLH instability. Thus, assuming $\omega=k_\parallel \rm{v_{b0}}+\delta=\omega_{\rm{dc}}+\delta$ (for DC instability) and $\omega=k_\parallel \rm{v_{b0}}+\delta=\omega_{\rm{dl}}+\delta$ (for DLH instability) separately, where $\delta~(\ll\omega_0,~\omega_{\rm{dc}},~\omega_{\rm{dl}})$ is the small frequency mismatch associated with the instabilities, we obtain the following expressions for the maximum growth rates $(\Im\delta)$.
\begin{equation}\label{eq-growth-dc1-max}
\gamma^{\max}_{\rm{dc}} \approx \frac{\sqrt{3}}{2}\left[\frac12\frac{\omega_{\rm{dc}}\left(\omega_{\rm{dc}}^{2}-\Omega_{\rm{d}}^{2}\right)}{\omega_{\rm{dc}}^{2}-\omega_{\rm{dl}}^{2}}\left(1+\frac{\alpha_b}{\eta}\right)\frac{\omega_{\rm{pb}}^{2}k^2_\parallel\lambda^2_{\rm{D}}}{1+k^2_\parallel\lambda^2_{\rm{D}}} \right]^{1/3},  
\end{equation}
\begin{equation}\label{eq-growth-dl1-max}
 \gamma^{\max}_{\rm{dl}} \approx \frac{\sqrt{3}}{2}\left[\frac12\frac{\omega_{\rm{dl}}\left(\omega_{\rm{dl}}^{2}-\Omega_{\rm{d}}^{2}\right)}{\omega_{\rm{dl}}^{2}-\omega_{\rm{dc}}^{2}}\left(1+\frac{\alpha_b}{\eta}\right)\frac{\omega_{\rm{pb}}^{2}k^2_\parallel\lambda^2_{\rm{D}}}{1+k^2_\parallel\lambda^2_{\rm{D}}} \right]^{1/3}. 
\end{equation}
From Eqs. \eqref{eq-growth-dc1-max} and \eqref{eq-growth-dl1-max}, we find that the maximum growth rates remain finite, but they gradually increase with increasing values of $k_\parallel\lambda_{\rm{D}}$ and eventually saturate at higher values of the wave number.   
Physically, as the wave number increases significantly, the dispersion properties of the wave change, shifting the system away from optimal resonance and thereby effectively limiting the growth of instability. However, the specific mechanism of saturation at large wave numbers is typically complex and depends heavily on the local plasma parameters (e.g., the coupling parameter $Z_e$ and the dust cyclotron frequency of the magnetic field strength ($\Omega_d$). Note that wave numbers much larger than unity may not be admissible. Otherwise, the plasma collective behaviors will be lost, as the wave number is normalized by the inverse of the effective Debye length. 
 The saturation of such instability growth rates may occur in various plasma environments, including planetary magnetospheres and rings, such as Saturn's rings, where, unlike the present model, the instability is driven by ion drag force or streaming particles with collisional effects. Although there is no direct observation, the observed  range of instabilities in laboratory experiments is consistent with these mechanisms.  
\par 
On the other hand, in the beam-cyclotron interactions, we consider the resonances at $\omega=k_\parallel \rm{v_{b0}}-\Omega_b$ and $\omega=\omega_{\rm{dc}}$ for the DC instability and $\omega=k_\parallel \rm{v_{b0}}-\Omega_b$ and $\omega=\omega_{\rm{dl}}$ for the DLH instability, and neglect the terms proportional to $k^2_\parallel$ $(\ll k_\perp^2)$ in Eq. \eqref{eq-DR2}.  Thus, assuming $\omega=k_\parallel \rm{v_{b0}}-\Omega_b+\delta=\omega_{\rm{dc}}+\delta$ and $\omega=k_\parallel \rm{v_{b0}}-\Omega_b+\delta=\omega_{\rm{dl}}+\delta$ separately, where $\delta~(\ll\omega_0,~\omega_{\rm{dc}},~\omega_{\rm{dl}})$  is the small frequency mismatch associated with the instabilities, we obtain the following growth rates of instabilities $(\Im\delta)$ corresponding to  DC and DLH modes.
\begin{equation}\label{eq-growth-dc2}
\gamma^{\max}_{\rm{dc}} \approx \frac{1}{2}\left[\frac{\omega_{\rm{dc}}\left(\omega_{\rm{dc}}^{2}-\Omega_{\rm{d}}^{2}\right)}{\Omega_b\left(\omega_{\rm{dc}}^{2}-\omega_{\rm{dl}}^{2}\right)}\left(1+\frac{\alpha_b}{\eta}\right)\frac{\omega_{\rm{pb}}^{2}k^2_\perp\lambda^2_{\rm{D}}}{1+k^2_\perp\lambda^2_{\rm{D}}} \right]^{1/2},  
\end{equation}
\begin{equation}\label{eq-growth-dl2}
 \gamma^{\max}_{\rm{dl}} \approx \frac{1}{2}\left[\frac{\omega_{\rm{dl}}\left(\omega_{\rm{dl}}^{2}-\Omega_{\rm{d}}^{2}\right)}{\Omega_b\left(\omega_{\rm{dl}}^{2}-\omega_{\rm{dc}}^{2}\right)}\left(1+\frac{\alpha_b}{\eta}\right)\frac{\omega_{\rm{pb}}^{2}k^2_\perp\lambda^2_{\rm{D}}}{1+k^2_\perp\lambda^2_{\rm{D}}} \right]^{1/2}. 
\end{equation} 
Similar to Cerenkov interactions, the maximum growth rates remain finite, and they tend to increase but saturate at higher values of $k_\parallel\lambda_{\rm{D}}$. However, in contrast to Eqs. \eqref{eq-growth-dc1-max} and \eqref{eq-growth-dl1-max}, the growth rates in Eqs. \eqref{eq-growth-dc2} and \eqref{eq-growth-dl2} are inversely proportional to the beam-cyclotron frequency, indicating that the higher (lower) the beam cyclotron frequency, the lower (higher) the growth rate. Thus, the maximum growth rates in beam-cyclotron interactions may be lower in magnitude compared to the Cerenkov interactions, implying that the Cerenkov resonant interaction gives rise to more efficient energy transfer from the beam to the wave compared to the beam-cyclotron resonance.
\section{Results and discussions} \label{sec-results}
\begin{figure*}[!h]
\centering
\includegraphics[width=15cm]{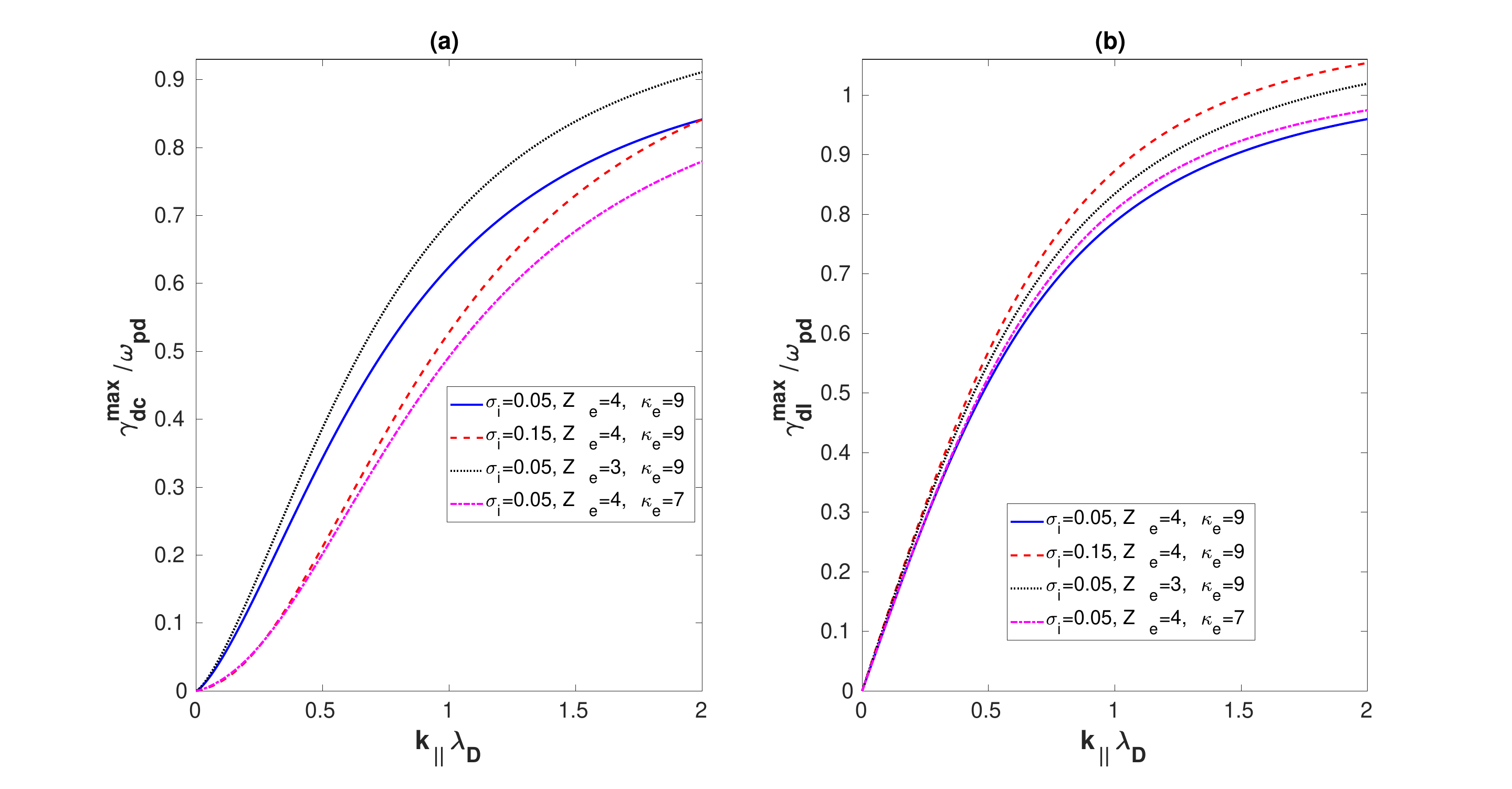}
\caption{In Cerenkov interactions, the profiles of the maximum growth rates (normalized by $\omega_{\rm{pd}}$) are shown against the parallel component of the wave number (normalized by $\lambda_{\rm{D}}^{-1}$) for different values of the parameters $\sigma_i$, $Z_e$, and $\kappa_e$ as in the legends. Subplots (a) and (b) correspond to instabilities associated with dust-cyclotron [Eq. \eqref{eq-growth-dc1-max}] and dust-lower-hybrid [Eq. \eqref{eq-growth-dl1-max}] modes respectively. The other fixed parameter values are $T_e/T_b=100$, $\rm{v}_{\rm{te}}/\rm{v}_{\rm{b0}}=0.01$, $\delta_e=2.5$, $\delta_i=1.4$, $\kappa_i=3$, $r_{d}/\lambda_{\rm{De}}=1.6$, and $\Omega_{\rm{dp}}\equiv\Omega_d/\omega_{\rm{pd}}=2$. }
\label{fig_1}
\end{figure*}

\begin{figure*}[!h]
\centering
\includegraphics[width=15cm]{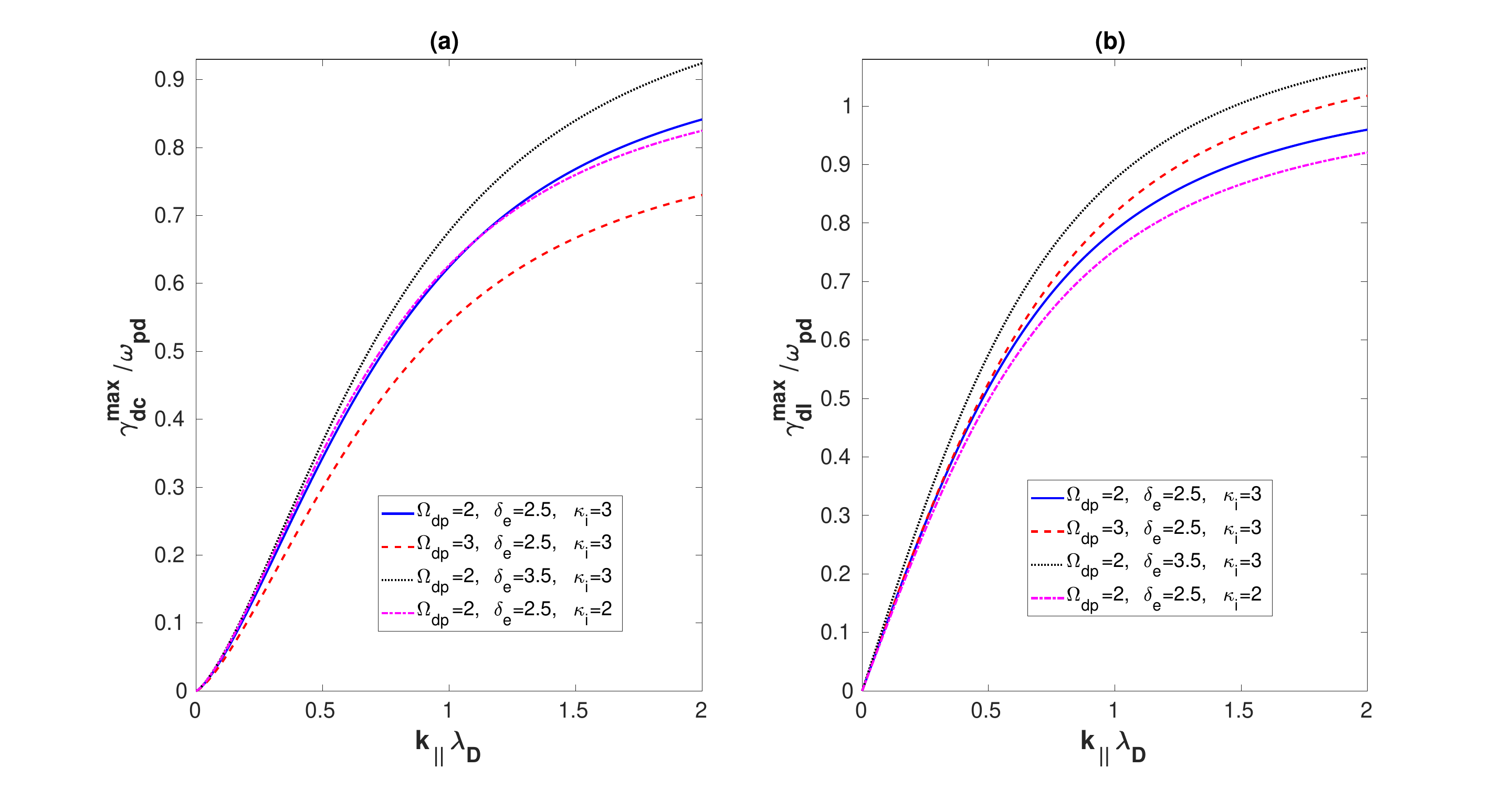}
\caption{The same as in Fig. \ref{fig_1} but with the variation of the parameters $\Omega_{\rm{dp}}\equiv\Omega_d/\omega_{\rm{pd}}$, $\delta_e$, and $\kappa_i$ as in the legends. The other fixed parameter values are $T_e/T_b=100$, $\rm{v_{te}}/\rm{v_{b0}}=0.01$, $\sigma_i=0.05$, $\delta_i=1.4$, $\kappa_e=9$, $r_{d}/\lambda_{\rm{De}}=1.6$, and $Z_e=4$.}
\label{fig_2}
\end{figure*}

\begin{figure*}[!h]
\centering
\includegraphics[width=15cm]{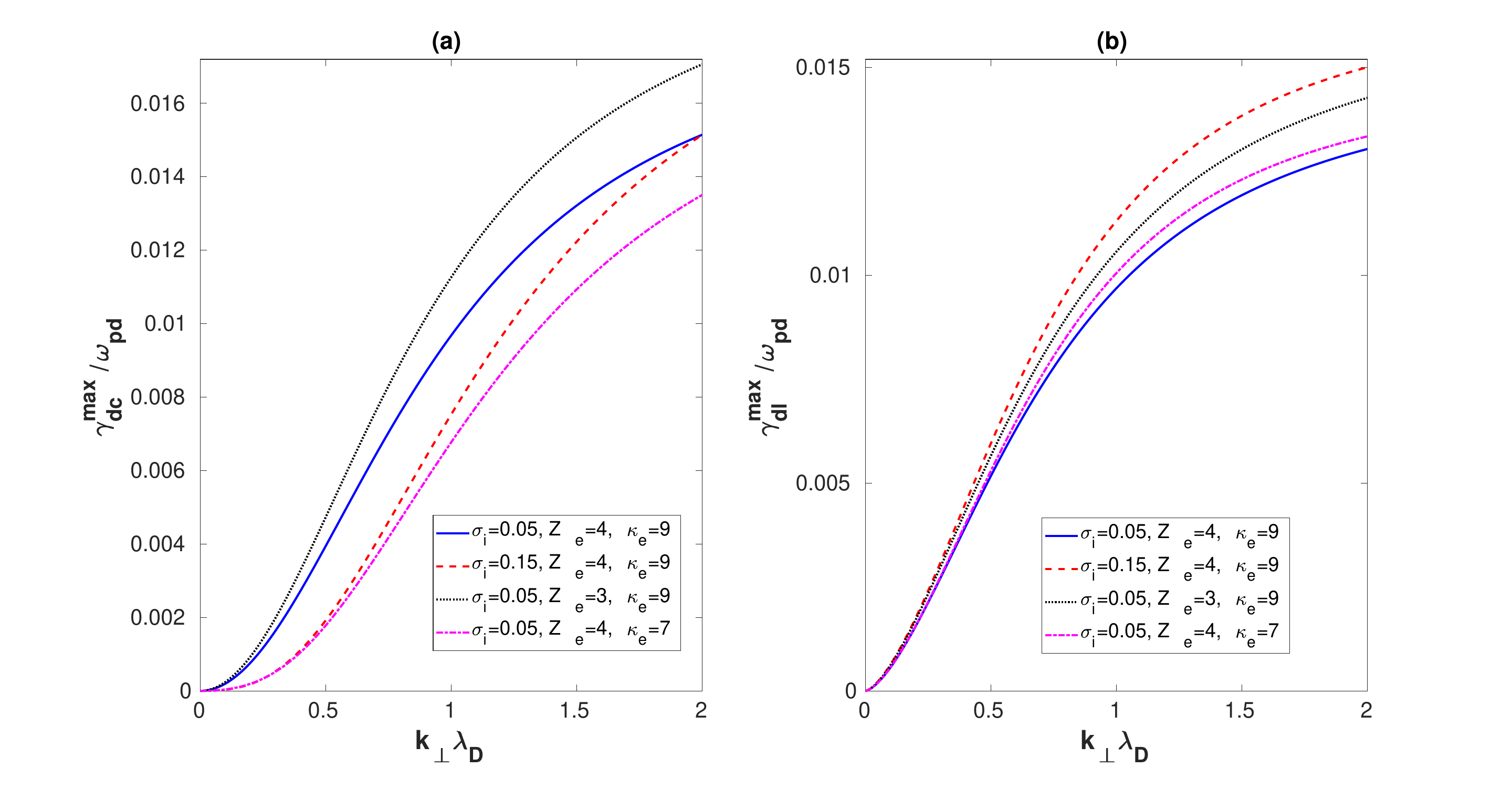}
\caption{In beam-cyclotron interactions, the profiles of the maximum growth rates (normalized by $\omega_{\rm{pd}}$) are shown against the perpendicular component of the wave number (normalized by $\lambda_{\rm{D}}^{-1}$) for different values of the parameters $\sigma_i$, $Z_e$, and $\kappa_e$ as in the legends. Subplots (a) and (b) correspond to instabilities associated with dust-cyclotron [Eq. \eqref{eq-growth-dc2}] and dust-lower-hybrid [Eq. \eqref{eq-growth-dl2}] modes respectively. The other fixed parameter values are the same as in Fig. \ref{fig_1}. }
\label{fig_3}
\end{figure*}

\begin{figure*}[!h]
\centering
\includegraphics[width=15cm]{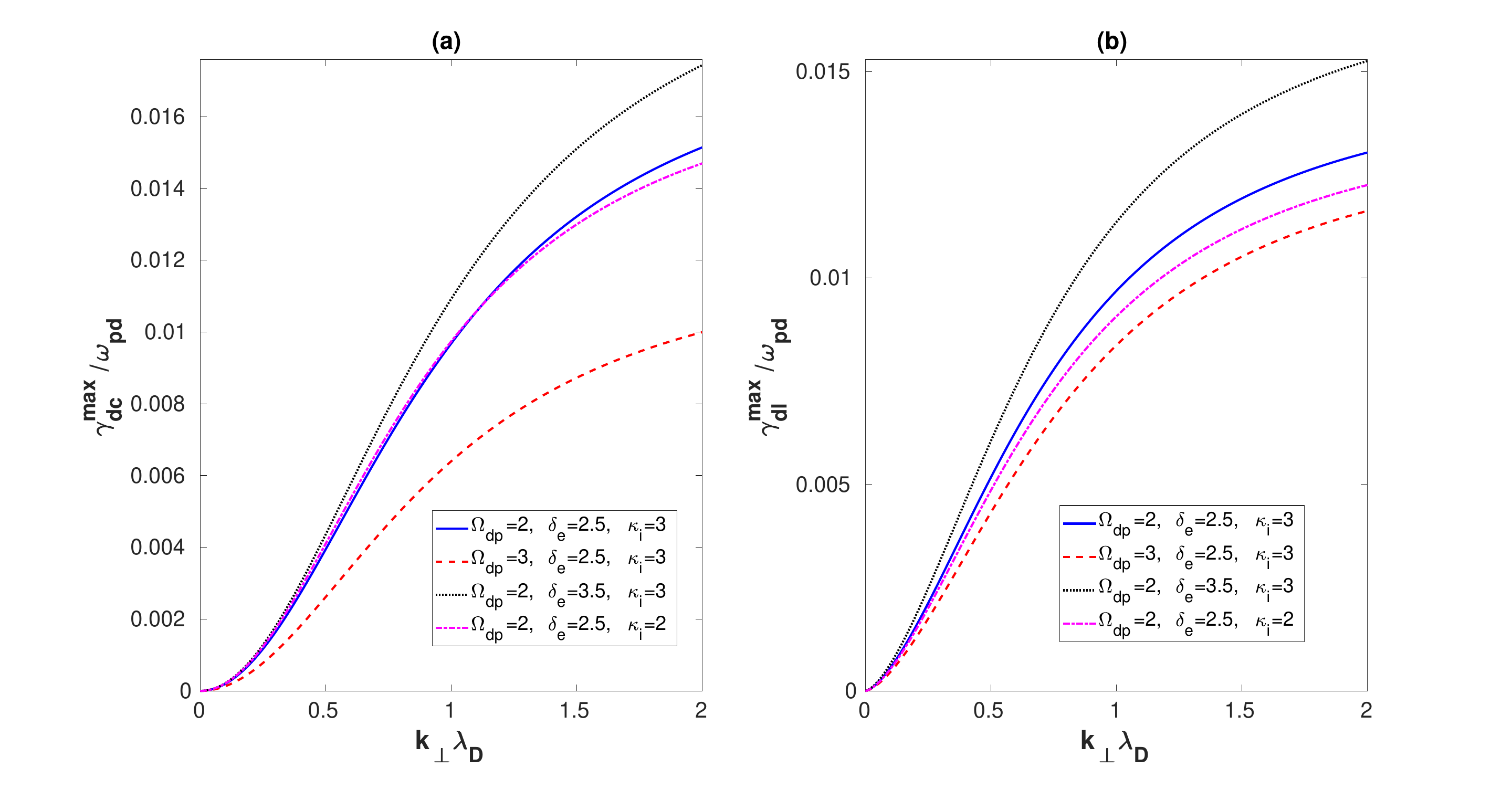}
\caption{The same as in Fig. \ref{fig_3} but with the variation of the parameters $\Omega_{\rm{dp}}\equiv\Omega_d/\omega_{\rm{pd}}$, $\delta_e$, and $\kappa_i$ as in the legends. The other fixed parameter values are $T_e/T_b=100$, $\rm{v_{te}}/\rm{v_{b0}}=0.01$, $\sigma_i=0.05$, $\delta_i=1.4$, $\kappa_e=9$, $r_{d}/\lambda_{\rm{De}}=1.6$, and $Z_e=4$.}
\label{fig_4}
\end{figure*}

We numerically analyze the growth rates of instabilities of DC and DLH waves as given by Eqs. \eqref{eq-growth-dc1-max} and \eqref{eq-growth-dl1-max} due to Cerenkov interactions and Eqs. \eqref{eq-growth-dc2} and \eqref{eq-growth-dl2} for beam-cyclotron interactions. To this end, we consider typical plasma parameters that are relevant to the Earth's magnetosphere \cite{popel2011effect,ma1998formation,mendis1994cosmic}, and define the dimensionless parameters as $\delta_{\textrm{j}} = n_{\textrm{j}0}/Z_{\textrm{d}0}n_{\textrm{d}0}$ (j=e, i, b) with $\delta_{\textrm{b}} =\delta_{\textrm{e}} - \delta_{\textrm{i}} + 1$, $\sigma_{\textrm{i}}=T_{\textrm{i}}/T_{\textrm{e}}$, $T_{\textrm{eb}} = T_{\textrm{e}}/T_{\textrm{b}}$, $V_{\textrm{tb}} = \text{v}_{\textrm{te}}/\text{v}_{\textrm{b}0}$, $Z_{\textrm{e}} = Z_{\textrm{d}}e^{2}/r_{\textrm{d}}T_{\textrm{e}}$,  $Z_{\textrm{i}} = Z_{\textrm{e}}/\sigma_{\textrm{i}}$, $Z_{\textrm{b}} = Z_{\textrm{e}}T_{\textrm{eb}}V_{\textrm{tb}}^{2}$, $\Omega_{\rm{dp}}\equiv\Omega_d/\omega_{\rm{pd}}$, and $\Omega_{\rm{bp}}\equiv\Omega_b/\omega_{\rm{pd}}$. Figures \ref{fig_1} and \ref{fig_2} show the profiles of the maximum growth rates of DC [Subplot (a)] and DLH [Subplot (b)] waves due to Cerenkov interactions, while the growth rates due to the beam-cyclotron interactions are displayed in Figs. \ref{fig_3} and \ref{fig_4} with the variations of the parameters as in the legends.  
\par 
Inspecting on the DC and DLH frequencies ($\omega_{\textrm{dc}}$ and $\omega_{\textrm{dl}}$), the charging coefficient $\alpha_{\textrm{b}}$, and the expression for $\widetilde{\alpha}_{\textrm{c}}$, we find that on changing the various parameters such as $\sigma_{\textrm{i}}$, $Z_{\textrm{e}}$, and $\kappa_{\textrm{e}}$, the DC wave frequency $\omega_{\textrm{dc}}$ is found to change while the DLH frequency $\omega_{\textrm{dl}}$ remains the same. Further, the charging coefficient for beam $\alpha_{\textrm{b}}$ remains the same for changing the parameters $\sigma_{\textrm{i}}$, and $\kappa_{\textrm{e}}$, while its values can increase by reducing the values of $Z_{\textrm{e}}$. However, the values of the ratio $\alpha_{\textrm{b}}/\eta$ increase with increasing values of $\sigma_{\textrm{i}}$. Also, the coefficient $\widetilde{\alpha}_{\textrm{c}}/\eta$ increases as one reduces the values of $Z_{\textrm{e}}$, which has a direct relation with $\omega_{\textrm{dc}}$ as evident from Eq. \eqref{eq-dc2}. These changes have noticeable impacts on the characteristics of the growth rates of DC and DLH modes. Figure \ref{fig_1} shows that while the maximum growth rate decreases for the DC mode with an increase in the ion to electron temperature ratio $\sigma_i$ [See subplot (a) and compare the solid and dashed lines)], the same for the DLH mode increases [See subplot (b) and compare the solid and dashed lines)]. It follows that the higher the temperature ratio, the higher the thermal motion of ions that eventually reduces the efficiency of resonant Cerenkov interactions for DC modes, but potentially increases for the DLH mode that drives the instabilities. Comparing the solid and dotted lines, we find that the instability growth rates for both DC and DLH modes get enhanced with a decrease in the coupling parameter $Z_e$ from $Z_e=4$ to $Z_e=3$. Physically, a reduction in $Z_e$ indicates weak electrostatic interactions between charged dust particles, leading to more fluid-like behaviors and hence enhanced growth rates. We also find that as the spectral index $\kappa_e$ gets lowered from $\kappa_e=9$ to $\kappa_e=7$, the DC growth rate ($\propto \omega_{\textrm{dc}}$) decreases but the DLH growth rate ($\propto \omega_{\textrm{dl}}$) gets enhanced [See the solid and dash-dotted lines in subplots (a) and (b)]. Physically, lower values of $\kappa_e$ correspond to more electrons with higher energies that significantly influence the Debye screening, the dust-charging process, and subsequent electrostatic interactions between charged dusts and plasma particles that drive the instabilities. An increased electron superthermal energy can also potentially alter the resonant interactions that contribute to the instabilities, leading to a decrease in the DC wave frequency $\omega_{\textrm{dc}}$ and an increase in the DLH wave frequency $\omega_{\textrm{dl}}$. 
\par
Next, we study the influences of the magnetic field strength, electron number density, and the ion spectral index, characterized by the dimensionless parameters $\Omega_d$,  $\delta_e$, and $\kappa_i$, on the profiles of the maximum instability growth rates, and show the results in Fig. \ref{fig_2}. The solid and dashed lines show that an increase in the dust-cyclotron frequency $\Omega_{\textrm{d}}$ results in a decrease (increase) in the DC (DLH) instability growth rate. For dust-cyclotron motions, an enhanced magnetic field strength restricts the motion of dust particles, influences them to gyrate about the magnetic field lines, thereby reducing them to participate in the resonant interaction, decreasing the DC wave frequency $\omega_{\textrm{dc}}$, and hence reduced growth rate [See the solid and dashed lines in subplot (a)]. On the other hand, the magnetic field modifies the DLH wave dispersion relation and the instability mode structure. In this case, the higher the magnetic field strength, the higher is the DLH wave frequency $\omega_{\textrm{dl}}$, which results in an enhanced growth rate [See the solid and dashed lines in subplot (b)]. The characteristics of the solid and dotted lines confirm that the growth rates get enhanced due to an increase in the electron concentration compared to the dust number density. Physically, an enhancement of the electron number density results in an increase in both the ion and beam number densities to maintain the charge quasineutrality. As a result, more particles will become involved in interacting with charged dust particles and contribute to the growth rates of instabilities of both DC and DLH modes. We note that the ion spectral index $\kappa_i$ does not significantly influence the DC instability growth rate; however, a decrease in $\kappa_i$ from $\kappa_i=3$ to $\kappa_i=2$ reduces the DLH growth rate (See the solid and dash-dotted lines). This occurrence is due to the significant mass difference between electrons and ions, in which superthermal ions are less energetic than superthermal electrons. Consequently, they do not significantly alter the interactions between dust grains and plasma particles, as well as the core dynamics responsible for the growth rate of instability associated with the dust's cyclotron motion. However, like superthermal electrons, more superthermal ions significantly alter the Debye screening length $\lambda_{\rm {D}}$, which in turn reduces the DLH wave frequency $\omega_{\textrm{dl}}$, thereby decreasing the instability growth rate.
\par
In the same way as for Cerenkov interactions [Figs. \ref{fig_1} and \ref{fig_2}], we show the profiles of the maximum growth rates of DC and DLH modes resulting from beam-cyclotron resonant interactions [Eqs. \eqref{eq-growth-dc2} and \eqref{eq-growth-dl2}] in Figs. \ref{fig_3} and \ref{fig_4} with the same sets of parameter values as for Cerenkov interactions. From Fig. \ref{fig_3}, we observe similar qualitative features of the growth rates with the variations of the parameters $\sigma_i$, $Z_e$, and $\kappa_e$, since the dependence of the growth rates on these parameters in both cases is the same. However, the key difference is that in contrast to Cerenkov interactions [Eqs. \eqref{eq-growth-dc1-max} and \eqref{eq-growth-dl1-max}], the growth rates in beam-cyclotron interactions depend explicitly on $\Omega_b$ and are inversely proportional to it, leading to reduced growth rates compared to Cerenkov interactions. Similarly, the effects of the variations of the electron number density ($\delta_e$) and ion spectral index ($\kappa_i$) will give rise to the same qualitative features of the growth rates for DC and DLH waves as for Cerenkov interactions. However, due to the mass difference between the charged dusts and ion beam, and the inverse relation of the growth rates with the beam cyclotron frequency ($\Omega_b\approx1000\Omega_d$), the growth rates for both DC and DLH waves get reduced with an increase in $\Omega_d$. Here, from Eq. \eqref{eq-growth-dl2}, we note that although the DLH wave frequency $\omega_{\rm{dl}}$ increases with an increase in $\Omega_d$, such an increase is not enough to intervene the reduction of the DLH growth rate by the effect of $\Omega_b$ or $\Omega_d$ (See Fig. \ref{fig_4}). In our numerical analysis, we have considered $\kappa$-parameter values relevant to the Earth's magnetosphere. In the latter, these values range from $3$ to $8$ for both electrons and ions. However, the magnetosphere is not uniform, so these values can vary depending on the region. We included a somewhat higher $\kappa$-value for electrons ($\kappa_e=9$) and a lower value for ions ($\kappa_i=2$). This was done to determine whether there is any significant change in the profiles of the growth rates. We observed no significant qualitative change in the growth rates for $\kappa$ values beyond the regime $3<\kappa_e,~\kappa_i<8$.  In this context, an important observation is that some parameters significantly affect DLH instability growth but not DC instabilities. For example, the influence of $\kappa_i$ on the DLH instability growth rate is more pronounced than the DC instability (See Fig. \ref{fig_2}). On the other hand, the parameter $\kappa_e$ has a greater impact on the DC instability growth rate than the DLH instability (See Fig. \ref{fig_1}). Typically, the DLH instability mechanism is more tied to bulk plasma properties and is thus more kinetically driven by the velocity distribution of ions than DC instability, making it inherently more sensitive to the presence and characteristics of superthermal ions. In contrast, the DC instability relies heavily on resonance conditions and is thus highly dependent on the electron velocity distribution, i.e., superthermal electrons in the high-energy tail, leading to a more pronounced influence of $\kappa_e$ on the growth rate. The effects of the other parameters on both the DLH and DC growth rates are seen to be significant. 
\section{Conclusions} \label{sec-conclu}
We have revealed a new dispersive dust-lower-hybrid-like mode and its coupling with the modified dust-cyclotron wave in a dusty magnetoplasma with dust-charge fluctuations. Such hybrid modes and the coupling were previously overlooked in the study of resonant cyclotron instabilities in a beam-driven magnetized dusty plasma \cite{prakash2014effect}. The mode propagates with a frequency lower than the typical dust-lower-hybrid frequency, $\widetilde{\omega}_{\rm{dl}}$. We have found that the streaming ion beam can couple the dust-lower-hybrid and dust-cyclotron modes, and instabilities of these modes occur due to both Cerenkov and beam-cyclotron resonant interactions that are associated with parallel and perpendicular wave propagation. In the absence of the ion beam, the modes become decoupled, and there will be no resonance and hence no instabilities. In this case, we recover the dust-cyclotron mode $\omega=\omega_{\rm{dc}}$, modified by the dust-charge fluctuation effect and the dust-lower-hybrid mode, $\omega=\omega_{\rm{dl}}$. The maximum instability growth rates for both the dust-lower-hybrid and dust-cyclotron modes appear to be higher in the Cerenkov interactions (and hence efficient energy transfer between the beam and waves) compared to the beam-cyclotron interactions. However, in both cases, they tend to reach steady states at higher values of the wave number. We have observed that the qualitative features of the instability growth rates in the Cerenkov and beam-cyclotron interactions are similar by the effects of the plasma parameters, namely the ion to electron temperature ratio $(\sigma_i)$, the Coulomb coupling parameter $(Z_e)$, the nonthermal parameters $(\kappa_e,~\kappa_i)$, and the electron to dust density ratio $(\delta_e)$ except the magnetic field effect on the dust-lower-hybrid instability growth rate. The latter gets increased by the influence of the magnetic field in the Cerenkov interaction, but it gets significantly reduced in the case of beam-cyclotron interactions.
\par 
Some important points are to be noted. In our analysis, we have assumed cold ion beams ($v_{\rm{tb}}\ll v_{\rm{b0}}$). The inclusion of a finite ion beam temperature will indeed influence the dust-charge fluctuation dynamics, and hence the plasma wave modes and the instabilities. Depending on the specific beam temperature and velocity regimes, the ion beam can introduce damping instead of instability. For example, when the beam temperature is comparable to or can exceed the electron temperature, the ion beam can lead to damping.  Such considerations require further study, but could be a project for future work. Also, we have considered ion-beam-driven resonant interactions, but neglected possible electron-beam effects. Since electrons are much lighter than ions, they typically reside far from the resonance condition with the low-frequency dust modes, namely, the DLH and DC modes, due to their much higher thermal and streaming velocities and shorter timescales. Consequently, for the specified typical dusty plasma conditions, the contributions of electron beams to the growth rates of the instabilities are not very noticeable. However, when the beam is extremely cold, dense, and slow enough that its streaming velocity equals the DLH or DC wave phase velocity, their effects may become comparable, and such conditions are rare in both laboratory and space dusty plasmas.
\par
 Some noteworthy observations are as follows:  
\begin{itemize} 
\item[•]  Increasing the ion temperature relative to the electron temperature results in a decrement of $\gamma^{\max}_{\rm{dc}}$ but enhances $\gamma^{\max}_{\rm{dl}}$.
\item[•] Both the maximum growth rates $\gamma^{\max}_{\rm{dc}}$ and $\gamma^{\max}_{\rm{dl}}$ can be enhanced by reducing the coupling strength $Z_e$.  
\item[•] As one approaches from superthermal to Maxwellian electron distributions with increasing values of $\kappa_e$, the maximum growth rates for dust-cyclotron modes  $(\gamma^{\max}_{\rm{dc}})$ significantly increase, while those  of dust-lower-hybrid modes $(\gamma^{\max}_{\rm{dl}})$ get reduced for both Cerenkov and beam-cyclotron interactions. On the other hand, the effects of superthermal ions on $(\gamma^{\max}_{\rm{dc}})$ are insignificant, but the values of $(\gamma^{\max}_{\rm{dl}})$ can be increased with increasing values of $\kappa_i$. 
\item[•] Increasing the magnetic field strength or the dust-cyclotron frequency results in significant reductions of the maximum growth rates except for $\gamma^{\max}_{\rm{dl}}$ associated with the Cerenkov interaction. 
\item[•] An enhancement of the electron density relative to the dust density $(\delta_e)$ increases the maximum growth rates for both Cerenkov and beam-cyclotron interactions.

\end{itemize} 
\par 
To conclude, since the dust-cyclotron and dust-lower-hybrid modes get coupled by the influence of the ion beam and become unstable due to Cerenkov and beam-cyclotron resonant interactions, the dust-cyclotron instability can drive the excitation of dust-lower-hybrid waves and vice versa. Such instabilities can enhance particle transport across the magnetic field lines. Such an enhancement can be significant in weakly coupled magnetized dusty plasmas with higher electron number densities. Furthermore, the energy from the instability can be transferred to electrons and ions, leading to plasma heating. These instabilities are relevant to various plasma environments, including space plasmas (e.g., Earth's magnetosphere) and laboratory dusty plasma experiments. In the latter, instabilities can derange plasma confinement by enhancing cross-field transport where the Cerenkov and cyclotron resonances between the streaming ion beam and the hybrid mode allow direct transfer of kinetic energy from the ion beam to both the DC and DLH waves.
Although lower-hybrid waves are commonly observed and extensively studied in the Earth's magnetosphere using spacecraft data (e.g., Van Allen Probes), as well as some other environments like planetary rings and cometary tails, and laboratory experiments have observed related phenomena like coupling between lower-hybrid waves and ion-cyclotron modes (See, e.g., Ref. \cite{kumar2022}), there is no published space observational evidence of the coupling between dust-cyclotron and dust-lower-hybrid modes.  In some spacecraft observations and laboratory experiments, the present theory is consistent with frequency ranges and coupling mechanisms observed in related low-frequency dusty plasma modes.
Finally, the coupling and instabilities reported here should be useful to investigate various nonlinear phenomena, such as the modulational instability of DLH wave envelopes coupled to the slow response from DC modes that are driven by the DLH wave ponderomotive force. The nonlinear evolution equations can be a coupled nonlinear Schr{\" o}dinger (NLS)  and a Korteweg-de Vries (KdV)-like equations. They are also be useful to investigate parametric excitations and the generation of a new kind of wave structure (e.g., oscillons). However, such an investigation is beyond the scope of the present work and left for future studies. 

\section*{ACKNOWLEDGMENTS}
NPA acknowledges research fellowship from the University Grants Commission, Bhaktapur, Nepal, with reference number PhD-78/79-S\&T-17. 
S. Basnet's research was supported by an appointment to the Young Scientist Training (YST) program at the APCTP through the Science and Technology Promotion Fund and Lottery Fund of the Korean Government, as well as, by the Korean Local Governments-Gyeongsangbuk-do Province and Pohang City.
\section*{Data availability statement}
All data that support the findings of this study are included within the article (and any supplementary files).
\section*{Author contributions}
\textbf{Amar Prasad Misra:} Model proposal; Conceptualization (lead); Formal analysis (equal);
Investigation (lead); Methodology (equal); Supervision (lead);
Validation (lead); Writing--original draft (equal); Writing – review \&
editing (equal). \textbf{Num Prasad Achary:} Conceptualization (supporting); Formal
analysis (equal); Investigation (equal); Methodology (equal);
Validation (supporting); Writing – original draft (equal).  \textbf{Suresh Basnet:}
Conceptualization (supporting); Formal analysis (supporting);
Investigation (equal); Methodology (supporting); Validation (supporting). \textbf{Raju Khanal:} Conceptualization (supporting); Formal analysis (supporting); Investigation (supporting); Methodology (supporting); Validation (supporting).
\section*{Conflict of interest}
The authors have no conflicts to disclose.
\bibliographystyle{apsrev4-2}
\bibliography{ref-rev}
\end{document}